\documentstyle[12pt]{article}
\newcommand{\be}{\begin{equation}}
\newcommand{\ee}{\end{equation}}
\newcommand{\bea}{\begin{eqnarray}}
\newcommand{\eea}{\end{eqnarray}}
\newcommand{\ba}{\begin{array}}
\newcommand{\ea}{\end{array}}

\def\bftau{\mbox{\boldmath $\tau$}}

\def\bftau{\mbox{\boldmath $\tau$}}

\begin{document}


\begin{titlepage}
\rightline{DAMTP-2006-81}
\rightline{ITFA-2006-38}
\rightline{hep-th/0610253}

\vfill

\begin{center}
\baselineskip=16pt
{\Large\bf Pseudo-Supersymetry and the Domain-Wall/Cosmology Correspondence}

\vskip 0.3cm
{\large {\sl }}
\vskip 10.mm
{\bf ~Kostas Skenderis$^{*,1}$ and  Paul K. Townsend$^{\dagger,2}$}
\vskip 1cm
{\small
$^*$
Institute for Theoretical Physics, \\
University of Amsterdam,\\
Valckenierstraat 65, 1018 XE Ansterdam,\\
The Netherlands\\
}
\vspace{6pt}
{\small
$^\dagger$
Department of Applied Mathematics and Theoretical Physics,\\
Centre for Mathematical Sciences, University of Cambridge,\\
Wilberforce Road, Cambridge, CB3 0WA, U.K.
}
\end{center}
\vfill

\par
\begin{center}
{\bf
ABSTRACT}
\end{center}
\begin{quote}

The correspondence between domain-wall and cosmological solutions of  gravity coupled to scalar fields is explained.  Any domain wall solution that admits a Killing spinor is shown to correspond 
to a cosmology that admits a pseudo-Killing spinor: whereas the Killing spinor obeys a Dirac-type equation with  hermitian `mass'-matrix, the corresponding pseudo-Killing spinor obeys a Dirac-type equation with a anti-hermitian  `mass'-matrix. We comment on some implications of (pseudo)supersymmetry.

\vfill
\vfill
\vfill

$^\star$ {To appear in proceedings of IRGAC 2006}

\vfill
 \hrule width 5.cm
\vskip
2.mm
{\small
\noindent $^1$ skenderi@science.uva.nl \\
\noindent $^2$ p.k.townsend@damtp.cam.ac.uk. 
\\
}
\end{quote}
\end{titlepage}
\setcounter{equation}{0}
\section{Introduction}

Domain wall solutions of supergravity theories, in spacetime dimension $D=d+1$, have been intensively studied in recent years because of their relevance to gauge theories via holographic renormalization.  
Initial studies concentrated on the case for which the the $D$-dimensional spacetime is foliated by $d$-dimensional Minkowski spaces;  in other words,  flat domain walls.  More recently, attention has been focused on curved domain walls, specifically those foliated by $d$-dimensional anti-de Sitter (adS) spacetimes, although domain wall solutions foliated by $d$-dimensional de Sitter (dS) spacetimes have also been considered; we shall refer to these as ``(a)dS-sliced'' domain walls.  In all these cases, the maximal symmetry of the `slices'  implies that only scalar fields are relevant to the solution, so the general low-energy Lagrangian density of interest takes the form
\be\label{Lstart}
{\cal L} = \sqrt{-\det g}\left[ R -\frac{1}{2}|\partial\Phi|^2  - V(\Phi)\right]\, , 
\ee
for metric $g$, with scalar curvature $R$, and scalar fields $\Phi$ taking values in some Riemannian target space and with potential energy function $V$. 

In the supergravity context, models of this type arise as consistent truncations, and a solution for which the supersymmetry variation of all fermion fields vanish for non-zero supersymmetry spinor parameter $\epsilon$ is said to be  ``supersymmetric''. The vanishing of the supersymmetry variation of the gravitino field leads to a ``Killing spinor''  equation of the form ${\cal D}\epsilon=0$, where ${\cal D}$ is an exterior covariant derivative on spinors constructed from the standard spin connection and a ``superpotential'',  which is (generically) a multi-component function of the scalar fields that  determines the potential $V$ through a simple derivative formula. For domain wall solutions it turns out that the constraints on $\epsilon$ implied by the vanishing  supersymmetry variations of other fermion fields are algebraic integrability conditions for the differential Killing spinor equation, so they yield nothing new. Thus, supersymmetric domain wall solutions are those for which ${\cal D}\epsilon=0$ can be solved for some non-zero spinor $\epsilon$, which is called a ``Killing spinor''. 

One reason for interest in supersymmetric solutions of a supergravity theory, in particular supersymmetric domain wall solutions, is that supersymmetry implies stability; in  particular, it implies classical stability. However,  classical stability cannot depend on the fermionic field content; instead, it depends  only on the existence of  a Killing spinor. This  is a weaker condition than supersymmetry since, for example, the existence of a Killing spinor places no restriction on the spacetime dimension $D$.  This suggests a weaker definition of supersymmetry, which has become known as ``fake'' supersymmetry, according to which a solution is considered ``supersymmetric''  if it admits a Killing spinor.  However,  the concept  of  fake supersymmetry depends on an understanding of what constitutes a Killing spinor outside the supergravity context. There is no general definition, as far as we are aware, but if we restrict our attention to domain wall solutions then the problem can be reduced, essentially, to a specification of the restrictions to be imposed on the superpotential used to define the covariant derivative operator ${\cal D}$.  As we shall see, there is an ambiguity in this supergravity-independent definition of a Killing spinor, and one of  our purposes here is to explain how this ambiguity may be exploited in the context of  cosmology. 

In cosmology, the requirement of homogeneity and isotropy implies,  just as for domain walls,  that the only relevant fields other than the metric tensor are scalar fields,   so the Lagrangian density (\ref{Lstart}) 
also provides a general starting point for the study of Friedmann-Lema\^itre-Robertson-Walker (FLRW) cosmologies. In fact, there is a correspondence between domain-wall solutions and FLRW cosmologies
for these models. For every domain-wall solution of the model with scalar potential  $V$ there is a cosmological solution of the same model but with scalar potential $-V$, and {\sl vice-versa} \cite{Skenderis:2006jq}. Here we present the details of the analytic continuations that connect 
the domain-wall and cosmological spacetimes that are paired by this  ``Domain-Wall/Cosmology  correspondence''. Special cases have been noted  on many previous occasions but the generality of the correspondence seems not to have been hitherto appreciated.  It raises the question of how  special features of domain walls, such as supersymmetry, are to be interpreted in the context of cosmology. This point was addressed briefly in \cite{Skenderis:2006jq}: cosmologies that correspond to supersymmetric domain walls are ``pseudo-supersymmetric'' in the sense that they admit a ``pseudo-Killing'' spinor.  The possibility of pseudo-Killing spinors arises precisely from the above-noted ambiguity in the extension of the notion of a Killing spinor to fake supersymmetry. Here we explain this point in more detail and 
discuss possible implications of pseudo-supersymmetry for cosmology.

\setcounter{equation}{0}
\section{Domain-Wall/Cosmology Correspondence}

The $D$-dimensional spacetime metric for a $d$-dimensional domain-wall of maximal symmetry can be put into the form
\be\label{DWmetric} 
ds^2_D = dz^2 + 
e^{2\beta\varphi}\left[ - \frac{d\tau^2}{1+ k \tau^2} + \tau^2 d\Omega_+^2\right]\, ,
\ee
where we have introduced the $D$-dependent constant
\be
\beta =1/\sqrt{2(D-1)(D-2)}\, , 
\ee
and  $d\Omega_+^2$ is an $SO(1,d-1)$-invariant metric on the unit radius $d$-dimensional hyperboloid; we may choose coordinates such that 
\be
d\Omega_+^2 = d\psi^2 + \sinh^2\psi\,  d\Omega_{d-2}^2\, . 
\ee
A domain-wall spacetime is therefore determined by a scale function $\varphi(z)$ 
and a constant $k$, which we may restrict  to the values $0,\pm1$, without 
loss of generality. The local geometry of the ($d$-dimensional) hypersufaces of constant $z$ is 
de Sitter for $k=1$, Minkowski for $k=0$, and anti-de Sitter for $k=-1$. For a given `fiducial' choice of $z$, these spacetimes can be viewed as the wall's ``worldvolume''; the coordinate $z$ is thus a measure of distance from this fiducial worldvolume.  In order to preserve the local 
isometries of the wall's worldvolume, the scalar fields $\Phi$ must be restricted to be functions of $z$ only.

Leaving aside domain walls for the moment, we turn to cosmology. The $D$-dimensional spacetime metric for an FLRW cosmology  has the form
\be\label{FLRWmetric} 
ds^2_D = -dt^2 + 
e^{2\beta\phi}\left[ \frac{ dr^2}{1 -  k r^2} + r^2 d\Omega_-^2\right]
\ee
where $d\Omega_-^2$ is an $SO(d)$-invariant metric on the unit radius $d$-sphere; we may choose coordinates such that
\be
d\Omega_-^2 = d\theta^2 + \sin^2\theta \, d\Omega_{d-2}^2\, . 
\ee
FLRW cosmologies are therefore determined  by the scale function $\phi(t)$ and the constant $k$, which we may again restrict  to the values $0,\pm1$, without loss of generality. 
The ($d$-dimensional) constant $t$ hypersurfaces are spheres for $k=1$, Euclidean spaces for $k=0$ and hyperboloids for $k=-1$, corresponding to closed, flat and open FLRW universes, respectively. 
 In order to preserve homogeneity and isotropy, the scalar fields $\Phi$ must be restricted to be functions of $t$ only.

The above domain-wall and cosmological spacetimes are related by analytic continuation. To see this, we start from the domain-wall spacetime of (\ref{DWmetric}),  define the new variables 
\be
(t, r, \theta )=-i( z, \tau , \psi )\, , 
\ee
and then analytically continue to real values of $(t,r,\theta)$. This yields the FLRW metric (\ref{FLRWmetric}) if we define
\be
\phi(t) = \varphi(it)\, . 
\ee
This makes it appear that $\phi$ is a complex function of $t$ but it is actually a real function that solves the field equations of the model with opposite sign of both $V$ and $k$. A simple example is ${\rm adS}_D$ sliced by ${\rm adS}_d$, $d$-dimensional Minkowski, or ${\rm dS}_d$ spacetimes, which become the $k=1,0,-1$ representations of ${\rm dS}_D$ as FLRW universes. 
For solutions that involve the scalar fields $\Phi$, one must similarly reinterpret the functions
$\Phi(z)$ of the domain-wall spacetime as real functions of $t$, which we then rename 
(in a slight abuse of notation) as $\Phi(t)$. The reason that this analytic continuation always works,
in the sense that the real functions determining a domain-wall solution become real functions determining a cosmological solution, can be seen as follows \cite{Skenderis:2006jq}. 

Let us consider the domain-wall and cosmological solutions together by introducing a sign $\eta$ such that $\eta=1$ for domain walls and $\eta=-1$  for cosmologies. Then, in either case,  the metric  can be put in the form
\be\label{ansatz} 
ds^2_D = \eta \left(e^{\alpha\varphi} f \right)^2 dz^2 + 
e^{2\beta\varphi}\left[ - \frac{\eta\, d\tau^2}{1+ \eta k \tau^2} + \tau^2 d\Omega_\eta^2\right]\, ,  
\ee
where
\be
\alpha = (D-1) \beta = \sqrt{\frac{D-1}{2(D-2)}}\, , 
\ee 
and, in order to maintain $z$-reparametrization invariance, we have made the replacement $dz \to e^{\alpha\varphi(z)}f(z)dz$ for   (lapse) function $f$, which must be monotonic  but is otherwise arbitrary; the gauge choice $f= e^{-\alpha\varphi}$ yields the forms of the domain-wall or cosmological metrics given above. The scalar fields $\Phi$ are functions only of $z$, which is a space coordinate for $\eta=1$ and a time coordinate for $\eta=-1$. The Euler-Lagrange equations of (\ref{Lstart})  then reduce to equations for the variables $(\varphi,\Phi)$ that are equivalent to the Euler-Lagrange  equations of the effective Lagrangian
\be\label{efflag}
L= \frac{1}{2}f^{-1} \left(\dot\varphi^2 -
|\dot\Phi|^2\right) 
- f e^{2\alpha\varphi}\left(\eta V(\Phi)
- \frac{\eta k}{2 \beta^2} e^{-2\beta\varphi}\right) \, , 
\ee
where the overdot indicates differentiation with respect to $z$.  It follows immediately from the form of this effective Lagrangian that {\it for every solution of the $\eta=1$ equations  of motion for potential $V$  there is a corresponding solution of the $\eta=-1$ equations of motion with potential $-V$,  with the opposite sign of $k$ if  $k\ne0$, and {\sl vice-versa}}. The domain-wall and cosmological solutions paired in this way are precisely those related by the analytic continuation procedure described above.

\setcounter{equation}{0}
\section{Fake Supersymmetry}

For a single scalar $\sigma$, the effective Lagrangian (\ref{efflag}) reduces to
\be\label{efflag2}
L= \frac{1}{2} f^{-1} \left(\dot\varphi^2 -
\dot\sigma^2\right) - f \, e^{2\alpha\varphi}\left(\eta V(\sigma)
- \frac{\eta k}{2 \beta^2} e^{-2\beta\varphi}\right) \, .  
\ee
It was observed in \cite{Celi:2004st} that the choice of target spaces coordinates can always be adapted to any given solution in such a way that this solution involves only a single scalar field, so the restriction to a single scalar is much less severe than one might  suppose. For this reason, many properties of single-scalar solutions can be extended to multi-scalar solutions. However, there are subtleties that arise in the application of this idea that we do not wish to enter into here, so we restrict ourselves to the one-scalar case. 

An example of a supergravity model with a single scalar field is the pure minimal $D=5$ gauged supergravity, and this provides a convenient, as well as physically relevant and historically significant, starting point for a study of fake supersymmetry of domain walls. The superpotential of this model is a real  $SU(2)$ triplet ${\bf W}$ and a straightforward generalization of the Killing spinor equation for this model suggests that we choose the exterior covariant derivative ${\cal D}$,  
mentioned in the introduction,  to be 
\cite{LopesCardoso:2001rt}
\be\label{calD}
{\cal D}= dx^\mu \left[D_\mu +  \alpha\beta \, {\bf W}\cdot \bftau\, \Gamma_\mu \right]\, , 
\ee
where $D_\mu$ is the standard  covariant derivative operator acting on Dirac spinors, $\bftau$ is the triplet of Pauli matrices acting on $SU(2)$ spinors, and $\Gamma_\mu $ are the 
spacetime Dirac matrices.  By this definition, ${\cal D}$ acts on $SU(2)$ doublets of Lorentz spinors. These would satisfy a symplectic reality condition in the context of minimal $D=5$ supergravity but, in the spirit of fake supersymmetry, we relax this condition here. The factor of $\alpha\beta$ arises from a choice of normalization of ${\bf W}$; this is fixed by the relation between ${\bf W}$ and $V$, which in our conventions is
 \be\label{Vee}
V= 2\left[|{\bf W}'|^2 - \alpha^2|{\bf W}|^2\right]\, , 
\ee
where the prime indicates a derivative with respect to  $\sigma$.

For a domain-wall metric of the form (\ref{DWmetric}) (i.e.
for  gauge choice $f=e^{-\alpha \varphi}$), the Killing spinor 
equation implies 
\be\label{killing1}
\partial_z \epsilon = \alpha\beta \, {\bf W}\cdot \bftau \, 
\Gamma_{\underline z} \, \epsilon \, ,
\ee
where $\Gamma_{\underline z}$ is a {\it constant} matrix that squares to the identity, and 
\be\label{killing2}
\hat D\,  \epsilon = e^{\beta\varphi}\hat\Gamma \left[
\left(\beta/2\right)\, \dot\varphi   \, 
\Gamma_{\underline z} + \alpha\beta\, 
{\bf W} \cdot\bftau \right]\epsilon\, , \nonumber 
\ee
where $\hat D$ is the standard worldvolume exterior covariant derivative on spinors and  
$\hat\Gamma$  is the {\it worldvolume} Dirac matrix valued 1-form.  

The integrability conditions for these equations were discussed in detail in \cite{Freedman:2003ax} and a simplified analysis was presented 
in \cite{Skenderis:2006jq}. We will not repeat the full analysis here, but  
we note that 
(\ref{killing2}) has the integrability condition 
\be\label{int1}
\dot\varphi^2 = 4\alpha^2 |{\bf W}|^2 + (k/\beta^2) e^{-2\beta\varphi}\, .
\ee
which upon use of the field equations leads to
\be \label{sigmadot}
\dot{\sigma} = \pm 2 |{\bf W}'|.
\ee
The joint integrability condition of (\ref{killing1}) and 
(\ref{killing2}) is
\be\label{dilatinovar}
\left(\dot\sigma + 2{\bf W}' \cdot\bftau \Gamma_{\underline z}\right)\epsilon=0\, ,
\ee
which can be interpreted in the supergravity context as the condition 
arising from the vanishing of the supersymmetry variation of the 
super-partner of $\sigma$.  This and (\ref{sigmadot}) then lead to the 
projection equation
\be\label{proj}
\left(1\pm \Gamma\right) \epsilon =0 \, ,\qquad 
\Gamma = \frac{{\bf W}'\cdot\bftau}{|{\bf W}'|}\,
\Gamma_{\underline z}\, ,
\ee
so the domain wall is half supersymmetric.
In general, this equation has its own integrability condition, since 
$\Gamma$ is a function of $z$, and this implies that
\be\label{consist1}
\left({\bf W}'{}' + \alpha\beta {\bf W} \right) \times {\bf W}'=0\, . 
\ee
From this we deduce that ${\bf W}$ must take the form 
\be\label{WtoZ}
{\bf W} = {\bf n}\,  {\cal R}e\, Z (\sigma) + {\bf m}\, {\cal I}m\, Z(\sigma)\, , 
\ee
where ${\bf n}, {\bf m}$ are two orthonormal 3-vectors and $Z$ is a complex function, with $Z=W$ for real scalar function $W$ when $k=0$. 
In addition, a complete analysis of consistency requires
\be\label{consist2}
|{\bf W}\times {\bf W}'|^2 =-k (D-2)^2e^{-2\beta\varphi}|{\bf W}'|^2\, . 
\ee

It was shown in  \cite{Skenderis:2006jq}, by direct  construction of $Z(\sigma)$,  that 
any $k=0$ or $k=-1$ domain-wall solution admits a Killing spinor provided that  the function $\sigma(z)$ is strictly monotonic; all such solutions are therefore (fake) supersymmetric. The $k=0$ case is especially simple, and was discussed earlier  in \cite{Freedman:2003ax,Sonner:2005sj}.  A Hamiltonian perspective on the general construction may be found in \cite{Skenderis:2006rr}. A solution for which $\dot\sigma(z)$ has isolated zeros can be considered ``piecewise supersymmetric'' but the construction breaks down completely if the zeros of $\dot\sigma(z)$ accumulate. As shown in \cite{Skenderis:2006rr},
unstable adS vacua are accumulation points and hence domain wall spacetimes that are asymptotic to an unstable adS vacuum are not (fake) supersymmetric, as expected since they are also unstable. 
If we agree, for the sake of simplicity,  to leave aside these exceptions, 
then we can summarize the 
result  by saying that {\it all flat or adS-sliced walls are 
(fake) supersymmetric}. 

This result was obtained  for a particular choice of operator ${\cal D}$, so we should consider 
to what extent  it depends on this choice.  In the supergravity context, the superpotential is generally in some non-trivial representation of the R-symmetry group that acts on the gravitino field, this being  $SU(2)$ for minimal $D=5$ supergravity. We took this $D=5$ example, with real $SU(2)$-triplet  superpotential, as our starting point and generalized it to arbitrary dimension $D$, relaxing the symplectic reality condition on the spinor in the process. Recall that one of the implications of integrability is that the real triplet superpotential is actually determined by a complex function
$Z$. The restrictions on $Z$ imposed by (\ref{consist1}) and (\ref{consist2}) could have been found directly by taking $D=4$ minimal supergravity as the starting point because the superpotential in this case is naturally a complex function; this route is less convenient, however, because of the complications of chirality. Note that an attempt to further simplify by assuming a real singlet superpotential (as suggested by $D=3$ minimal supergravity) would restrict  fake supersymmetry to flat domain walls. This is an unnecessary restriction,  so it is important to consider whether some more general superpotential might similarly show that the restrictions obtained by the assumption of a real-triplet superpotential are similarly unnecessary. For example, one could consider \cite{Zagermann:2004ac} the $USp(4)$ 5-plet superpotentials  suggested by extended $D=5$ supergravity. 
It would be surprising if this were to allow new 
possibilities\footnote{Note however that such more general superpotentials 
may be useful in establishing the existence of more than one fake 
supersymmetry.} because the real triplet superpotential 
is already unnecessarily general; it is determined by a complex function. Moreover, the triplet superpotential is already sufficient to establish the fake supersymmetry of almost all flat or adS-sliced domain walls. There are good physical reasons for the exceptions, as noted above. Nor should we expect to discover that some more general superpotential that will allow the dS-sliced walls to be considered fake-supersymmetric  because (for $D>2$) there is no physically 
acceptable supersymmetric extension of the de Sitter group.
These considerations fall short of being a proof but they convince us that domain-walls that are not (fake) supersymmetric in the particular sense described above will not become (fake) supersymmetric for some other choice of the exterior differential operator ${\cal D}$.

\section{Pseudo-supersymmetry}

As already observed, in certain spacetime dimensions it may be possible to impose a symplectic reality  condition on $\epsilon$, and this is required for $D=5$ minimal supergravity since the minimal $D=5$ spinor is an ``$SU(2)$-Majorana''  spinor.  A symplectic reality condition on $\epsilon$ effectively enforces the reality of ${\bf W}$ since complex conjugation of the Killing spinor equation ${\cal D}\epsilon=0$ then yields the same equation but with ${\bf W}$ replaced by its complex conjugate. 
Once the symplectic reality condition on $\epsilon$ is relaxed, there is no immediate reason why ${\bf W}$ should be real, although the reality of $V$ implies that it must be either real or pure imaginary. Of course, if ${\bf W}$ is pure imaginary, then we can redefine it to be real at the cost of changing the covariant derivative ${\cal D}$ from the expression given in (\ref{calD}) to
\be\label{calD2}
{\cal D} = dx^\mu\left(D_\mu +  i\alpha\beta \, {\bf W}\cdot \bftau\, \Gamma_\mu\right)\, . 
\ee
At the same time, we must change the relation (\ref{Vee}) to
\be\label{Vee2}
V= -2\left[|{\bf W}'|^2 - \alpha^2|{\bf W}|^2\right]\, .  
\ee
This can be viewed as the same relation as (\ref{Vee})  but for a model with  scalar potential of opposite sign from the original model. 

At this point we see how the ambiguity in the notion of a Killing spinor outside the supergravity context might be exploited in cosmology, because the cosmological `dual' of a domain-wall solution of a given model with scalar potential $V$ is a solution of the same model but with $V$ replaced by $-V$. For the FLRW metric (\ref{FLRWmetric}) the use of (\ref{calD2}) yields the equations
\be\label{killing3}
\partial_t \epsilon = \alpha\beta \, {\bf W}\cdot \bftau \, 
i\Gamma_{\underline t} \, \epsilon \, ,
\ee
where $\Gamma_{\underline t}$ is a {\it constant} matrix that squares to {\it minus} the identity, and 
\be\label{killing4}
\hat D_a \epsilon = e^{\beta\varphi}i\hat\Gamma_a\left[
-\left(\beta/2\right)\, \dot\varphi   \, 
i\Gamma_{\underline t} + \alpha\beta\, 
{\bf W} \cdot\bftau \right]\epsilon\, , \nonumber 
\ee
where $\Gamma_a$ are the Dirac matrices, in a coordinate basis, for a fiducial spacelike hypersurface of fixed $t$.  This has the integrabity condition
\be\label{intcos}
\dot\varphi^2 = 4\alpha^2 |{\bf W}|^2 - (k/\beta^2) e^{-2\beta\varphi}\, . 
\ee
This is the same as (\ref{int1}) if we take $k\to -k$, which is a consequence of the overall factor of $i$ on the right hand side of (\ref{killing4}). The joint integrability conditions of (\ref{killing3}) and (\ref{killing4}) 
is
\be
\left(\dot\sigma + 2{\bf W}' \cdot\bftau i\Gamma_{\underline t}\right)\epsilon=0\, .
\ee
Because $i\Gamma_{\underline t}$ squares to the identity, this is 
equivalent to 
(\ref{dilatinovar}) and hence leads to an equivalent constraint on $\epsilon$ and, for $k\ne0$, equivalent constraints on ${\bf W}$. 
To summarize, given a domain-wall solution with $k=0,-1$ there is a construction of a real triplet superpotential  ${\bf W}$ such that the wall admits a Killing spinor. The corresponding $k=0,1$ cosmological solution of the model with $V\to -V$ then admits a spinor satisfying a similar equation but with ${\bf W} \to i{\bf W}$. We shall call such a spinor a ``pseudo-Killing'' spinor. The difference 
between Killing and pseudo-Killing spinors can be characterized as follows: by taking the `gamma-trace'  of the (pseudo)-Killing spinor equation we deduce that the (pseudo-)Killing spinor satisfies a Dirac type equation of the form
\be
{\cal D}\!\!\!\!/ \, \epsilon = M \epsilon\, , 
\ee
where $M$ is a `mass' matrix (albeit a non-constant one). For a genuine Killing spinor this mass matrix 
is hermitian whereas for a pseudo-Killing spinor it is anti-hermitian. 

We have now seen how to construct a pseudo-Killing spinor for a cosmological solution of a model of the type defined by (\ref{Lstart}) starting from a Killing spinor associated to a supersymmetric domain wall solution of the same model but with opposite sign potential.  Given that almost all flat or adS-sliced domain wall solutions are supersymmetric, we may now conclude that almost all flat or closed FLRW cosmologies are `pseudo-supersymmetric'.  Perhaps the simplest, although very special, example of a supersymmetric domain wall is a stable adS vacuum. The dS spacetime that is its  cosmological `dual'  is then pseudo-supersymmetric.  In the following section we shall explore some implications of this fact.

\setcounter{equation}{0}
\section{Applications}

Anti de Sitter space can be viewed as a special case of a flat domain wall spacetime. It can also be viewed as either an adS-sliced or a dS-sliced domain wall, but the standard Minkowski slicing will be sufficient for present purposes. Given a potential $V(\sigma)$, maximally symmetric vacua correspond 
to  constant values of $\sigma$ for which $V(\sigma)$ is extremized.  Let us suppose that $V$ has an extremum at $\sigma=0$ with $\eta V_0 <0$, so that  the vacuum is $adS_D$ for  $\eta=1$ and $dS_D$ for $\eta=-1$.  In this case $V$ has the Taylor expansion
\be\label{taylor}
 V= -{\eta\over 2\beta^2 \ell^2} + {1\over2}  m^2 \sigma^2 + {\cal O}\left(\sigma^3\right)
\ee
where $\ell$ is the (a)dS radius and $m$ the mass of the scalar field fluctuation. Let $V$ be given by 
\be\label{VW}
\eta V= 2\left[ \left(W'\right)^2 - \alpha^2W^2\right]\, , 
\ee
for real singlet superpotential $W$. This is just (\ref{Vee}) with ${\bf W}$ given by (\ref{WtoZ}) with 
$Z=W$. As mentioned earlier, a real singlet superpotential suffices for consideration of flat 
domain walls, and can be found for any $V$, at least in principle,  by solving the differential equation  
(\ref{VW}) for $W$.  We therefore have
\be\label{prebound}
\eta V' = 4W' \left(W'{}' -\alpha^2 W\right)\, , 
\ee
from which we see that there are two types of (a)dS vacua. Those for which $W'=0$ and those for which $W'\ne0$ (in which case $W'{}'= \alpha^2 W$).  The adS vacua with $W'=0$ are the supersymmetric 
vacua; this terminology is consistent with our earlier terminology for  domain walls because when $\sigma$ is constant and $\epsilon$ is non-zero, the supersymmetry preserving condition (\ref{dilatinovar}) reduces to $W'=0$. Similarly, the dS vacua with $W'=0$ are the pseudo-supersymmetric vacua. By differentiating (\ref{prebound}) and evaluating at the stationary point 
of $W$, one can derive a bound on $m^2$. For $\eta=1$, this is the Breitenlohner-Freedman (BF) bound \cite{Breitenlohner:1982bm,Mezincescu:1984ev}
\be\label{genBF}
m^2 \ge   - \frac{(D-1)^2}{4\ell^2}\, , 
\ee
which states that $m$ cannot be ``too tachyonic'',  and the method of proof is the one of  \cite{Boucher:1984yx,Townsend:1984iu}. The BF bound is not  absolute because it  is a trivial matter to construct a model with an adS vacuum that violates the bound; these vacua are not associated with stationary points of $W$ and are not supersymmetric. Nevertheless, adS vacua that satisfy the bound are physically distinct from those that do not; the former are stable, at least classically, whereas the latter 
are unstable. 

The same procedure leads, for $\eta=-1$ to a cosmological analog of the BF bound. This is the 
 {\it upper} bound \cite{Skenderis:2006rr}
\be
m^2\le \frac{(D-1)^2}{4\ell^2}\, . 
\ee
Again, the bound is not absolute but serves to separate dS vacua with distinct physical properties. 
If this inequality is satisfied then the scalar field $\sigma$ approaches its equilibrium value at the dS vacua monotonically, like an overdamped pendulum. If the potential rises `too steeply' from its (positive) minimum, such that the bound is violated, then $\sigma$ will overshoot its equilibrium value and then oscillate about it  {\sl ad infinitum} as it approaches this value, like an underdamped pendulum. This implies that a dS vacuum violating the bound is an accumulation point for zeros of $\dot\sigma$. The same is true of adS vacua that violate the BF bound and this is why domain walls that are asymptotic to unstable adS vacua fail to be (fake) supersymmetric. In the cosmological case, however, it is less clear that a violation of the bound implies an instability. One might expect the oscillations implied by a violation of the bound to cause a particle production that could reduce the potential energy of the dS vacuum. Instabilities of dS space due to particle production have been proposed \cite{Mottola:1984ar} but also opposed \cite{Allen:1985ux}. We will not attempt to review the 
current situation here; it appears from a recent analysis \cite{Das:2006wg} that the matter is still not completely resolved.

\bigskip
\noindent
{\bf Acknowledgements.} PKT thanks the EPSRC for financial support.  KS is supported by NWO via the Vernieuwingsimplus grant  ``Quantum gravity and particle physics''. We thank Ashok Das and Jaume Garriga for useful discussions.


\begin{thebibliography}{99}

\bibitem{Skenderis:2006jq}
  K.~Skenderis and P.~K.~Townsend,
  {\it Hidden supersymmetry of domain walls and cosmologies}, 
  Phys.\ Rev.\ Lett.\  {\bf 96} (2006) 191301
  [arXiv:hep-th/0602260].
  
\bibitem{Skenderis:2006rr}
  K.~Skenderis and P.~K.~Townsend,
  {\it Hamilton-Jacobi method for domain walls and cosmologies}, Phys. Rev. D
{\bf 74} (2006)  125008
 [ arXiv:hep-th/0609056].
  
\bibitem{Celi:2004st}
  A.~Celi, A.~Ceresole, G.~Dall'Agata, A.~Van Proeyen and M.~Zagermann,
  {\it On the fakeness of fake supergravity}, 
  Phys.\ Rev.\ D {\bf 71} (2005) 045009
  [arXiv:hep-th/0410126].
  
\bibitem{LopesCardoso:2001rt}
  G.~Lopes Cardoso, G.~Dall'Agata and D.~L\"ust,
  {\it Curved BPS domain wall solutions in five-dimensional  gauged
  supergravity}, 
  JHEP {\bf 0107} (2001) 026
  [arXiv:hep-th/0104156].

 
\bibitem{Freedman:2003ax}
  D.~Z.~Freedman, C.~Nu\~nez, M.~Schnabl and K.~Skenderis,
 {\it Fake supergravity and domain wall stability}, 
  Phys.\ Rev.\ D {\bf 69}, 104027 (2004)
  [arXiv:hep-th/0312055].
  
\bibitem{Sonner:2005sj}
  J.~Sonner and P.~K.~Townsend,
  {\it Dilaton domain walls and dynamical systems},
  Class.\ Quant.\ Grav.\  {\bf 23} (2006) 441
  [arXiv:hep-th/0510115].
  

\bibitem{Zagermann:2004ac}
 M.~Zagermann,
 {\it N = 4 fake supergravity}, 
Phys.\ Rev.\ D {\bf 71} (2005) 125007
  [arXiv:hep-th/0412081].

\bibitem{Breitenlohner:1982bm}
  P.~Breitenlohner and D.~Z.~Freedman,
 {\it Positive Energy In Anti-De Sitter Backgrounds And Gauged Extended
  Supergravity}, Phys.\ Lett.\ B {\bf 115} (1982) 197; 
 {\it Stability In Gauged Extended Supergravity}, 
  Annals Phys.\  {\bf 144} (1982) 249.
  
\bibitem{Mezincescu:1984ev}
  L.~Mezincescu and P.~K.~Townsend,
  {\it Stability At A Local Maximum In Higher Dimensional Anti-De Sitter Space And
  Applications To Supergravity}, 
  Annals Phys.\  {\bf 160}, 406 (1985).

  
\bibitem{Boucher:1984yx}
  W.~Boucher,
  {\it Positive Energy Without Supersymmetry}, 
  Nucl.\ Phys.\ B {\bf 242} (1984) 282.
  
\bibitem{Townsend:1984iu}
  P.~K.~Townsend,
  {\it Positive Energy And The Scalar Potential In Higher Dimensional
  (Super)Gravity Theories}, 
  Phys.\ Lett.\ B {\bf 148} (1984) 55.
  

\bibitem{Mottola:1984ar}
  E.~Mottola,
  {\it Particle Creation In De Sitter Space}, 
  Phys.\ Rev.\ D {\bf 31} (1985) 754.
  
\bibitem{Allen:1985ux}
  B.~Allen,
  {\it Vacuum States In De Sitter Space}, 
  Phys.\ Rev.\ D {\bf 32} (1985) 3136.
  
\bibitem{Das:2006wg}
  A.~Das and G.~V.~Dunne,
  {\it Large-order perturbation theory and de Sitter/anti de Sitter effective
  actions}, 
  Phys.\ Rev.\ D {\bf 74} (2006) 044029
  [arXiv:hep-th/0607168].
  
 
    

\end{thebibliography}
\end{document}